\documentclass[letterpaper]{article} 
\usepackage{aaai24}  
\usepackage{times}  
\usepackage{helvet}  
\usepackage{courier}  
\usepackage[hyphens]{url}  
\usepackage{graphicx} 
\usepackage{natbib}  
\usepackage{caption} 
\usepackage{algorithm}
\usepackage{algorithmic}

\usepackage{newfloat}
\usepackage{listings}
\DeclareCaptionStyle{ruled}{labelfont=normalfont,labelsep=colon,strut=off} 
\floatstyle{ruled}
\newfloat{listing}{tb}{lst}{}
\floatname{listing}{Listing}
\title{Parallel Ranking of Ads and Creatives in Real-Time Advertising Systems}
\author{
    Zhiguang Yang,
    Lu Wang, 
    Chun Gan,
    Liufang Sang\thanks{Corresponding author},
    Haoran Wang,
    Wenlong Chen,
    Jie He,
    Changping Peng,
    Zhangang Lin,
    Jingping Shao
}
\affiliations{
    JD.com\\
    zgyang1996@gmail.com, wanglu319@jd.com, ganchun1@jd.com, sangliufang@jd.com, wanghaoran35@jd.com, chenwenlong17@jd.com, hejie67@jd.com, pengchangping@jd.com, linzhangang@jd.com, shaojingping@jd.com
}

\usepackage{bibentry}

\begin{document}

\maketitle

\begin{abstract}
``Creativity is the heart and soul of advertising services'' \cite{mkt:96}. 
Effective creatives can create a win-win scenario: advertisers can reach target users and achieve marketing objectives more effectively, users can more quickly find products of interest, and platforms can generate more advertising revenue. 
With the advent of AI-Generated Content, advertisers now can produce vast amounts of creative content at a minimal cost. 
The current challenge lies in how advertising systems can select the most pertinent creative in real-time for each user personally. 
Existing methods typically perform serial ranking of ads or creatives, limiting the creative module in terms of both effectiveness and efficiency. 
In this paper, we propose for the first time a novel architecture for online parallel estimation of ads and creatives ranking, as well as the corresponding offline joint optimization model. 
The online architecture enables sophisticated personalized creative modeling while reducing overall latency. 
The offline joint model for CTR estimation allows mutual awareness and collaborative optimization between ads and creatives. 
Additionally, we optimize the offline evaluation metrics for the implicit feedback sorting task involved in ad creative ranking.
We conduct extensive experiments to compare ours with two state-of-the-art approaches. 
The results demonstrate the effectiveness of our approach in both offline evaluations and real-world advertising platforms online in terms of response time, CTR, and CPM. 
\end{abstract}

\section{Introduction}
\begin{figure}[t!]
\centering
\includegraphics[width=0.9\columnwidth]{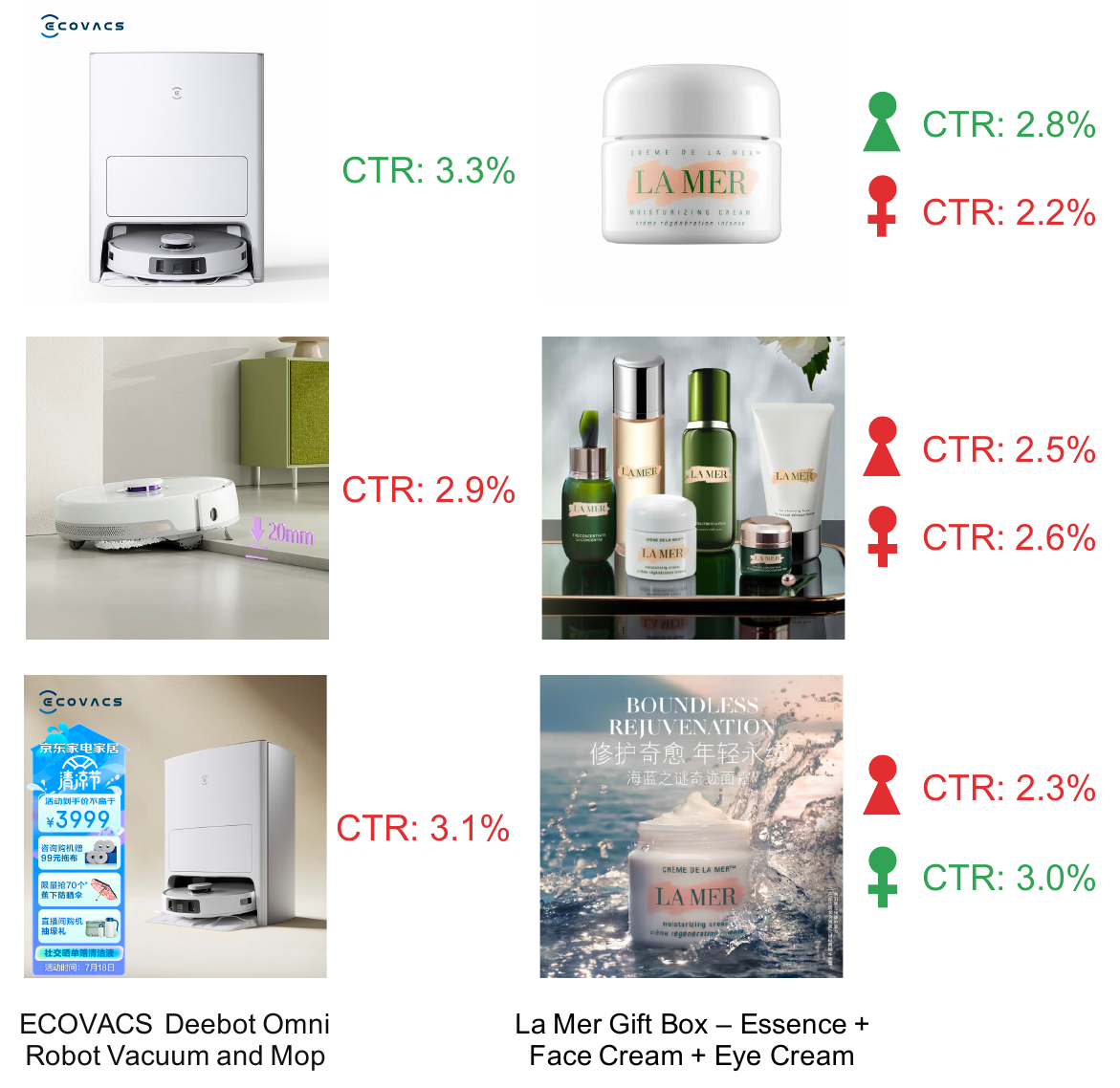}
\caption{The CTR comparison across different creatives for the same product, demonstrates that creatives can have a significant impact on CTR. 
Creatives can highlight product attributes in various ways, such as intuitively displaying the product itself, presenting scenarios of using the product, or emphasizing special pricing or other selling points of the product.
The first column shows the overall CTR for three creatives of the same product, underscoring that even well-designed creatives may underperform. 
The second column presents the CTR by gender for different creatives of the same product, further emphasizing the importance of personalized creative selection.}
\label{fig1}
\end{figure}

Online advertising has revolutionized the revenue generation landscape for e-commerce platforms such as Amazon, Taobao, and JD.com, emerging as a pivotal source of income in the digital era. 
These modern advertising platforms have harnessed the power of personalized advertising, leveraging user interests and preferences to deliver targeted product displays. 
Creatives, as shown in Fig.\ref{fig1}, lie at the heart of this advertising ecosystem, acting as a bridge between users and products. 
Through various display formats, such as images, titles, and videos, creatives spotlight the unique qualities and benefits of products to engage potential buyers. 
With the rapid evolution of AI-Generated Content(AIGC) technologies\cite{gpt3:20,sd:22}, advertisers now have the ability to generate an extensive range of creatives, tailored to different product attributes and styles, to captivate and engage users. 
Consequently, the creative ranking module, which can dynamically select the most compelling advertising creatives based on real-time user preferences has become increasingly crucial. 

Current recommendation systems mainly adopt a two-stage cascading architecture: candidate retrieval and ranking. 
In retrieval, billions of ad candidates are narrowed down to thousands. 
With M(M \textgreater 1000) ad candidates and N(N \textgreater 10) creative options per ad, the ranking stage needs to estimate M × N times to identify the optimal ads and creatives to display. 
However, under strict online time and computational constraints, simultaneous estimation is infeasible. 
Leading methods tackle this by sequentially ranking creatives and ads. 
The creative ranking model needs to rank an order of magnitude more items than the ad ranking within a less latency budget. 
Under strict time constraints, creative ranking requires highly efficient algorithms online, such as the bandit algorithm \cite{HBM:21,chen2021automated} and tree model \cite{chen2021efficient}. 
Recent studies have also highlighted the profound influence of creative ranking on the overall performance of ad systems. 
CACS\cite{CACS:22} upgraded from Post-CR architecture with bandit algorithms to Pre-CR with Deep Neural Networks (DNNs), based on a simplistic vector-product model. 
It consists of separate Query-User and Ad-Creative towers, with the final score calculated as the cosine distance between the two resulting vectors. 
This vector-product model has relatively weaker capabilities compared to approaches that first concatenate all features and feed them into a wide and deep fully-connected network\cite{wd:16}. 
However, the results reported by CACS showed that their method led to a 5.3\% increase in response time, but only a 3.1\% CTR improvement. 
The disproportionate growth in latency compared to incremental CTR gains indicates limitations of the Pre-CR architecture. 
More integrated architectures are needed to unleash the power of deep fully-connected networks for robust creative modeling without sacrificing efficiency.

In this work, we adopt a novel perspective on creative ranking by decoupling it from the conventional ad retrieval and ranking process. 
This strategy significantly reduces end-to-end latency and liberates creative ranking from previous latency constraints, thereby facilitating the use of more sophisticated models for enhanced click-through rate prediction. 
Additionally, to counteract the performance decline due to the ranking module's inability to perceive creative information, we train a large rank model offline. 
This model incorporates both ad and creative features. 
We have refined the model's architecture to allow for its division into two parallel online models. 
This division enables the ad model to account for biases introduced by the creative, while the creative model benefits from the ad model’s precise representations, thereby improving the accuracy of creative bias estimation.

The offline evaluation metrics for creative ranking remain an open problem. 
Existing works commonly use Area Under ROC (AUC)\cite{ROC:06} and GAUC\cite{gauc} to evaluate ad ranking performance. 
However, AUC is not suitable for evaluating creative ranking ability under the same ad. 
In industrial practice, the most used strategy is online A/B testing, but it impacts ad revenue and takes many days to produce a confident result. 
\cite{HBM:21} proposed simulated CTR (sCTR) to assess creative ranking ability, but it alters the sample distribution causing metric instability. 
We propose Normalized sCTR (NSCTR), which normalizes the sCTR calculation using the ad distribution in the sample. 
Through case studies and comparisons with online A/B tests, we demonstrate the efficacy of this new offline metric.

In summary, our contributions are as follows:

\begin{itemize}
    \item To the best of our knowledge, we are the first to perform parallel estimation of the creative ranking module and the ad ranking module named \textbf{Peri-CR}.
    This significantly reduces overall latency while increasing the time budget for each module to support more sophisticated and effective modeling. 
    \item We propose a joint optimization framework of ad and creative (\textbf{JAC}) to model the interactions and interdependencies between ads and creatives. 
    Through joint feedback training, we can achieve collaborative optimization for CTR estimation. 
    \item We optimize the offline evaluation metrics \textbf{NSCTR} for the implicit feedback task involved in creative ranking. This improves the model's capability to generalize to online performance. 
    We conduct extensive experiments on both offline and online advertising platforms. Results demonstrate the superiority of our approach over baselines in terms of CTR, CPM, and response time. 
\end{itemize}

\section{Related Work}
Our work is related to CTR prediction and ranking in advertising systems. 
In this section, we first briefly introduce related works on advertising ranking, followed by works focused on creative ranking. 

\textbf{Advertising Ranking. }
Modern advertising systems often use cascade ranking systems\cite{wd:16,mobius:19,cascade:17}  to select the most relevant items from billions of candidates. 
These cascade ranking systems mainly comprise two stages: recall and ranking. 
The recall stage retrieves tens of thousands from billions of candidates and feeds candidates into the ranking module\cite{prerank:23}. 
The ranking module then scores these candidates and outputs the top results to display to the user. 
Major research directions include feature extraction\cite{ItemSage:22,auto:22,alt:2023}, feature interaction\cite{dssm:13,wd:16,dcn:17,yang2023incremental}, and user behavior modeling\cite{din:18,dien:19,uniform:23,textseq:23}. 

\textbf{Creative Ranking. }
Creative ranking faces unique challenges compared to ad ranking. 
The creative pool is an order of magnitude larger while user feedback is far more sparse. 
This demands efficient and lightweight online estimation under tight time and space constraints. 
Thus, creative ranking research has focused more on offline assessments and designing high-efficiency online algorithms. 
Previous studies like NIMA\cite{nima:18} and PEAC\cite{peac:19} focused on offline creative quality evaluation based on image and text content. 
On the other hand, PEAC demonstrates the importance of online user feedback for creative ranking. 
Recent studies focus on creatives ranking online. 
AES\cite{aes:21} uses ingredient tree and Thompson sampling to select creatives. 
HBM-VAM\cite{HBM:21} shows the visual priors and a flexible updated bandit method can raise the platform revenue. 
CACS\cite{CACS:22} resembles our method the most and they place the creative ranking module before the ad ranking stage then jointly optimize them with distillation and share embedding. 

\begin{figure*}[t]
\centering
\includegraphics[width=0.8\textwidth]{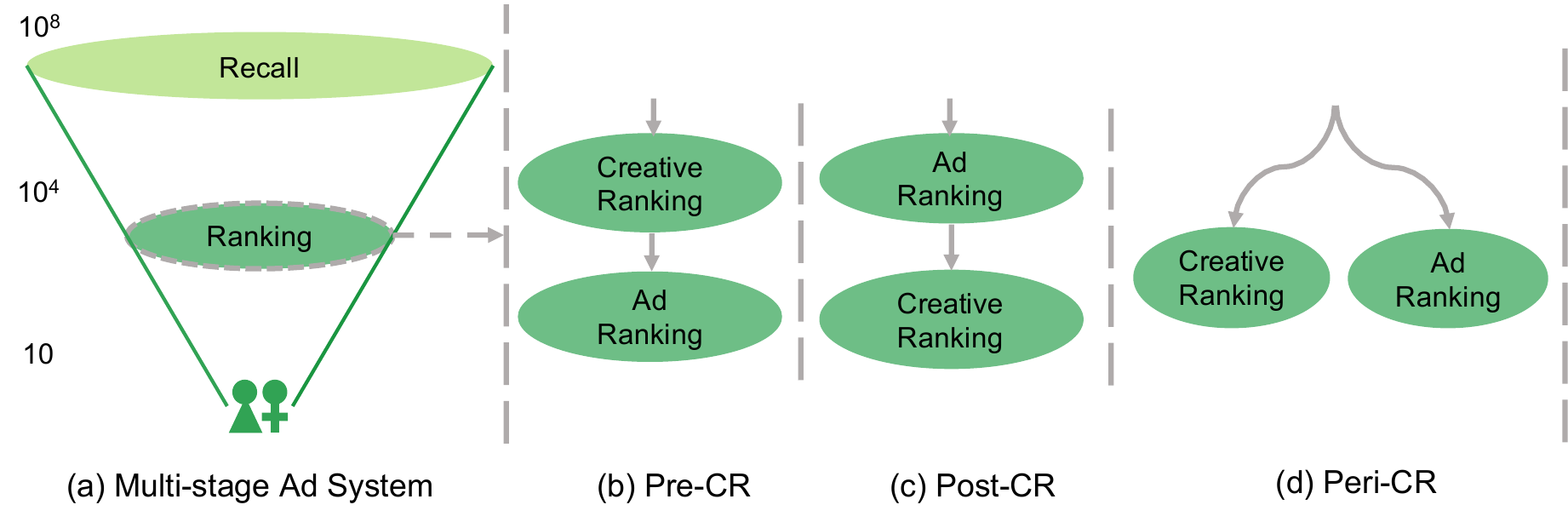} 
\caption{The main modules and workflow of the online advertising system. (a) Multi-stage ad ranking system. (b) Pre-CR first requests creative ranking then ad ranking. (c) Post-CR first requests ad ranking then creative ranking. (d) Peri-CR requests ad ranking and creative ranking in parallel simultaneously.}
\label{fig2}
\end{figure*}

Due to the strict latency constraints and system architecture limitations in modern advertising systems, existing works have to sacrifice precise creative estimation. 
While CACS\cite{CACS:22} achieves better system performance by adjusting the architecture, we further decouple the creative module from the main ad pipeline, opening up space for more accurate creative estimation.

\section{Problem Formulation}
An advertising platform tends to display advertisements with high eCPM(effective Cost Per Mille) to users, while eCPM is determined by the product of the CTR and CPC(Cost Per Click). 
$eCPM=CTR\cdot CPC\cdot 1000$. 
The scope of this work is constrained to ranking ads with the highest CTR from thousands of candidates, with less RT(response time).

The problem formulation is as follows. 
Given the set of Ad $a$ as $\{A_{1}, A_{2}, ..., A_{M}\}$, each Ads $A^{m}$ with a bundle of creative $c$, indicated as $\{C^{m}_{1}, C^{m}_{2}, ..., C^{m}_{N}\}$. 
The number of ad slots is $(L\ll M)$, they will be displayed to users $u$. 
The ranking models output the scores in each ad and creative and select the top $L$ ad and top $1$ creative according to the scores. 
Formally, given a triple $(u,a,c)$, the ranking model predicts the score $z$ as follows: 
\begin{equation}
    z=p(y=1|u,a,c) \label{1}
\end{equation}
where $p$ denotes the ranker model, and $y\in \{0,1\}$ is the label that denotes whether the user clicks the impressed ad. 
To simplify the formula notation, we will use $x$ to represent the user-ad pair $(u, a)$ in the following. 

The permutation of combination is unacceptable both in time and compatibility, we transform the formula:
\begin{equation}
    p(y=1|x) \cdot p(c|x,y=1) = p(c|x) \cdot p(y=1|x,c) \label{2}
\end{equation}

While the ranking model widely adopts the softmax function to predict the click probability: 
\begin{equation}
    p(c|x)=softmax(\frac{p(y|x,c)}{\sum\limits_{c_{i}^{'}\in C_N}{p(y|x,c=c_{i}^{'})}}) \label{3}
\end{equation}

Where $\sum\limits_{c_{i}^{'}\in C_N}{p(y|x,c=c_{i}^{'})}$ is an unknown constant. Combined with Equation\ref{2} and \ref{3},

\begin{equation}
f_\theta(x) \cdot e^{-f_\theta(x)} = p(y=1|x) \cdot p(c|x,y=1) \label{4}
\end{equation}
Where $f_\theta(x)$ denotes $p(y=1|x,c)$. 
Thus we decompose the permutation problem into the CTR prediction of ad and the creative distribution with the user and ad. 

\section{Methodology}

We propose Peri-CR, a novel architecture for creative and ad ranking, along with the corresponding model framework JAC, which improves the performance of both ranking tasks and reduces the overall response time.

\subsection{Online Parallelism Architecture}
For the ranking of ad and creative, there are three methodologies as shown in Fig.\ref{fig2}. 

\textbf{Post-CR} places the creative ranker after the ad ranker. 
It only ranks creative for the output of the ranking stage, so the number of ads that the creative model needs to evaluate is greatly reduced (less than 10) after truncation by the ad ranker. 
This requires less time and allows more precise modeling of creatives. 
However, the display creative for the ads cannot be determined during the ad ranking stage, which will damage the performance of this stage.

\textbf{Pre-CR}(CACS) put the creative ranker before the ad ranker. 
It can improve the performance of the ad ranker by considering the display creative but the creative ranker cost more time to evaluate creatives for all ad candidates. 
On the other hand, it also constrains the creative model to use a simple vector-product-based model.  
\begin{figure*}[t!]
\centering
\includegraphics[width=0.8\textwidth]{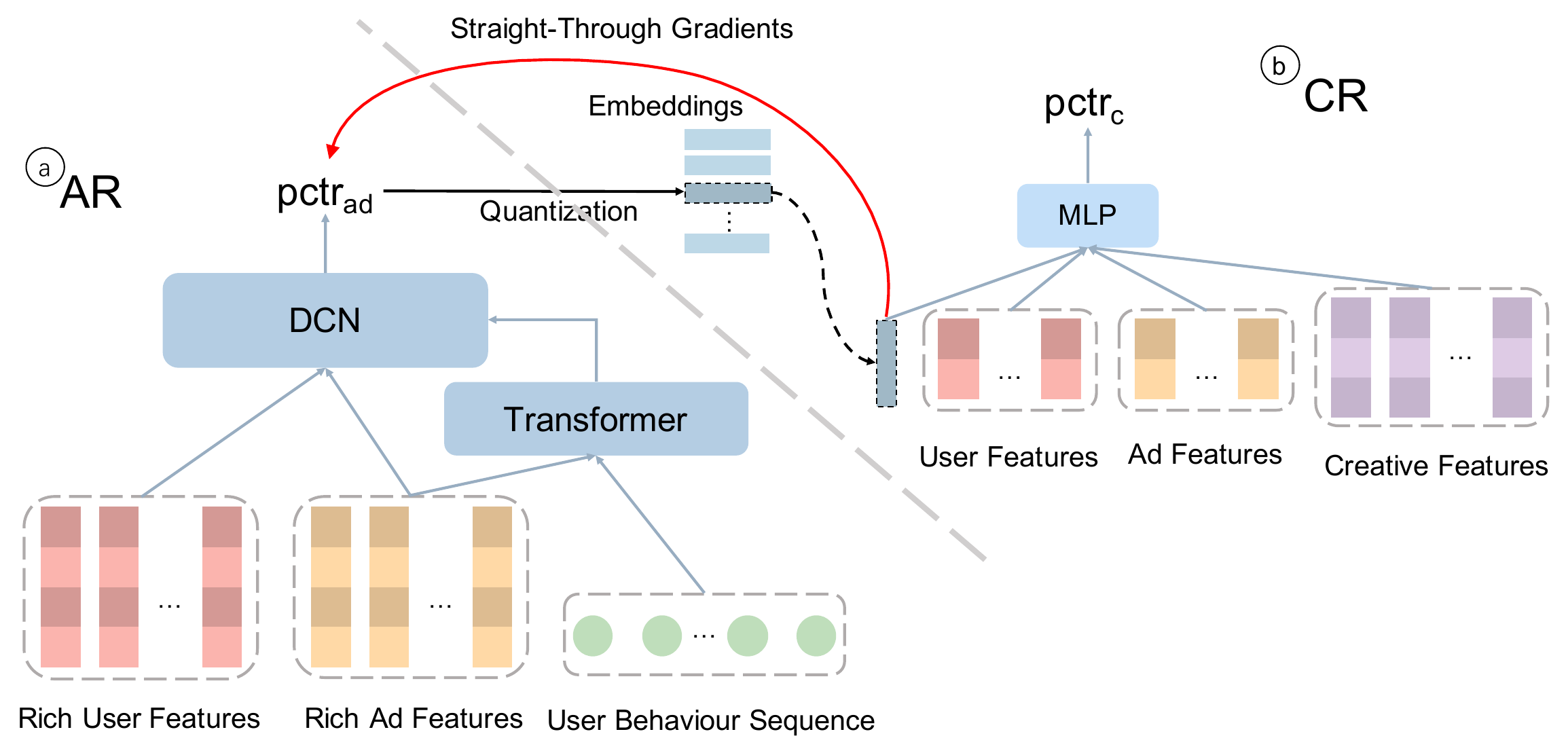} 
\caption{Framework of the proposed Joint optimization of Ad and Creative ranking(JAC), including two submodels: 
(a)Ad Ranking(AR) adopts deep cross network (DCN) as the main architecture to predict ad CTR, taking rich user and ad features as input. 
It also employs transformer to model user behavior sequence; 
(b)Creative Ranking(CR) uses a smaller network with fewer features, and leverages the AR output to estimate creative CTR.}
\label{fig3}
\end{figure*}

\textbf{Peri-CR}. 
Our proposed approach decouples the creative ranking module from the main ad retrieval and ranking pipeline. 
After the initial retrieval stage, the ad ranker and creative ranker are requested in parallel. 
The ad ranker takes abundant user and ad features as input and estimates the ad-level CTR $pctr_{ad}$. 
Concurrently, the creative ranker uses fewer user, ad, and creative features as input, and estimates the creative-level CTR $pctr_c$. 
The $pctr_c$ is used to determine the optimal creative to display for each ranked ad. 
And the $pctr_{ad}$ determines the final displayed ad sequence. 

This parallel decoupled architecture provides several advantages:
First, eliminating dependence between the ad and creative modules, allowing sufficient time budget for both to employ more sophisticated models. 
The ad ranker focuses solely on ad relevance, while the creative ranker specializes in creative appeal and diversity.
Second, separating the modules reduces overall latency compared to cascade selection. 

Although we can make ad ranker aware of creatives offline through our JAC model introduced next, it will still be unable to access accurate information about displayed creatives online.
One solution is incorporating creatives in a re-ranking stage\cite{rerank:19,rerank:21} if available. 
More importantly, there exists a trade-off between precise ad and creative ranking. 
\textit{"No silver bullet"}\cite{silver}. 
Systems should optimize overall performance by choosing suitable architectures based on their constraints. 

\subsection{Offline Joint Optimization} \label{jac}

A naive approach is to model and predict CTR separately with AR (Ad Ranking) and CR (Creative Ranking). However, this has several issues: 
(1) AR does not have access to creative information which is a strong bias that impacts CTR, hurting model performance. 
(2) CR needs to evaluate dozens of times more candidates compared to AR at the same time, requiring a simpler model which also leads to less accurate CTR estimates.

As Equation.\ref{2} we designed a large cascaded model offline that combines the ad ranking and creative ranking models. 
This ensures the two submodels are aware of each other, improving prediction accuracy, as shown in Fig.\ref{fig3}:

The left side is the original AR. Its inputs are the full user features, ad features, and cross features, performing complex user behavior sequence modeling. 
It outputs the CTR prediction of ad $pctr_{ad}$.

The right side is CR. Its primary inputs are simple user features, ad features, and rich creative features, passed through a small MLP network to output creative CTR estimates $pctr_c$. 
Another input feature is the output of AR $pctr_{ad}$. $pctr_{ad}$ goes through logarithmic transformation and quantization before lookup embedding, its quantized code is calculated as
\begin{equation}
    \lfloor K \cdot log_{r+1}(1+r\cdot pctr_{ad}) \rfloor \label{5}
\end{equation}
where $K$ is embedding size and $r$ is the hyperparameter that can be calculated by information gain. 

This embedding is initialized and concatenated with other feature embeddings as input to upper network layers. 
Since the quantization process (Equation.\ref{5}) is non-differentiable, the gradients cannot be directly propagated through it. 
To enable backpropagation to the AR, the gradients are directly copied from the CR to the AR\cite{VQVAE:17}. 
This allows CR to leverage rich information from AR for more accurate estimation while suppressing CTR estimate instability. 

During online inference, it can be easily split into two parallel modules to concurrently estimate. 
AR uses $pctr_{ad}$ as output. 
Since CR is responsible for estimating creatives, the $pctr_{ad}$ is the same for the same ad across a request, so CR directly uses the historical statistical CTR as input for each ad.

\subsection{Quality of the Implicit Sub-Ranking}
For the creative ranking model, we design NSCTR to measure the performance. 

Area Under ROC(AUC) is the most popular offline metric of the ranking model, which is computed over the whole exposure. 
While it brings user bias, GAUC can focus on the performance of the ad ranking list for each user. 
The creative ranker is applied to rank candidate creative for each user-ad pair. 
AUC and GAUC are often dominated by user-ad ranking. 
It is natural to consider GAUC grouped by user-ad. 
However, generally an ad has only one exposure per user, insufficient to compute GAUC. 

Simulated CTR(sCTR)\cite{HBM:21} is the metric designed for creative selection. 
It replays the recorded impression data offline, filters the record which the selected creative by the offline model differs from online exposed creative, and then accumulated their impressions and clicks. 
$sCTR = click/impression$. 
We find that the calculation of sCTR will change the distribution of the sample. 
The result tends to the CTR of the ad which has less creative candidates. 
Here is a simple example. 
$Ad_1$ has 2 creatives displayed randomly, with a CTR of 0.2. 
$Ad_2$ has 3 creatives displayed randomly, with a CTR of 0.3. 
The average $aCTR$ is 0.25. 
Offline we also randomly select creatives to display. 
When calculating sCTR, $Ad_1$ will filter out half of the samples, and $Ad_2$ will filter out two-thirds of the samples. 
So $sCTR=0.24 < aCTR$, which indicated that the offline random algorithm achieved better results than online, which is clearly unreasonable.

We present the NSCTR, calculated as  Algorithm.\ref{alg:algorithm}:

\begin{algorithm}[b!]
\caption{Evaluation Metrics - NSCTR}
\label{alg:algorithm}
\textbf{Input}: impression data $I$ with $A$d, $C$reative, and $y$(click or not), creative ranker $f_{cr}$\\
\textbf{Output}: NSCTR
\begin{algorithmic}[1] 
\STATE Let $impressions \leftarrow 0$;
\STATE Let $clicks \leftarrow 0$;
\STATE Let $\{ImpA_1,...,ImpA_M\} \leftarrow 0$;
\STATE Let $\{Imp^{s}A_1,...,Imp^{s}A_M\} \leftarrow 0$;
\STATE Let $\{Clk^{s}A_1,...,Clk^{s}A_M\} \leftarrow 0$;
\FORALL{ impression $\{(A_m,C_n,y)^{i}\}^{I}_{i=1}$}
\STATE $ImpA_m \leftarrow ImpA_m + 1$;
\FOR {$C_k$ in Creatives $\{C_1, ..., C_N\}$ Given in $A_m$ }
\STATE Get predicted scores $y_{k}=f_{cr}(A_m,C_k)$;
\ENDFOR
\STATE Choose the creative $C_{k'}$ = argmax($y_{1}$,...,$y_{K}$);
\IF {$C_{k'}=C_{n}$}
\STATE $Imp^{s}A_m \leftarrow Imp^{s}A_m + 1$;
\STATE $Clk^{s}A_m \leftarrow Clk^{s}A_m + y$;
\ENDIF
\ENDFOR
\STATE $impressions = \sum\{ImpA_1,...,ImpA_M\}$;
\STATE $clicks = \sum \{\frac{Clk^{s}A_1*ImpA_1}{Imp^{s}A_1},...,\frac{Clk^{s}A_M*ImpA_M}{Imp^{s}A_M}\}$;
\STATE $NSCTR=clicks/impressions$;
\STATE \textbf{return} $NSCTR$;
\end{algorithmic}
\end{algorithm}

We approximate the CTR for all samples by using the CTR of the samples where the top-1 creative selected by the offline model matches the creative exposed online.

\begin{figure}[t]
\centering
\includegraphics[width=0.9\columnwidth]{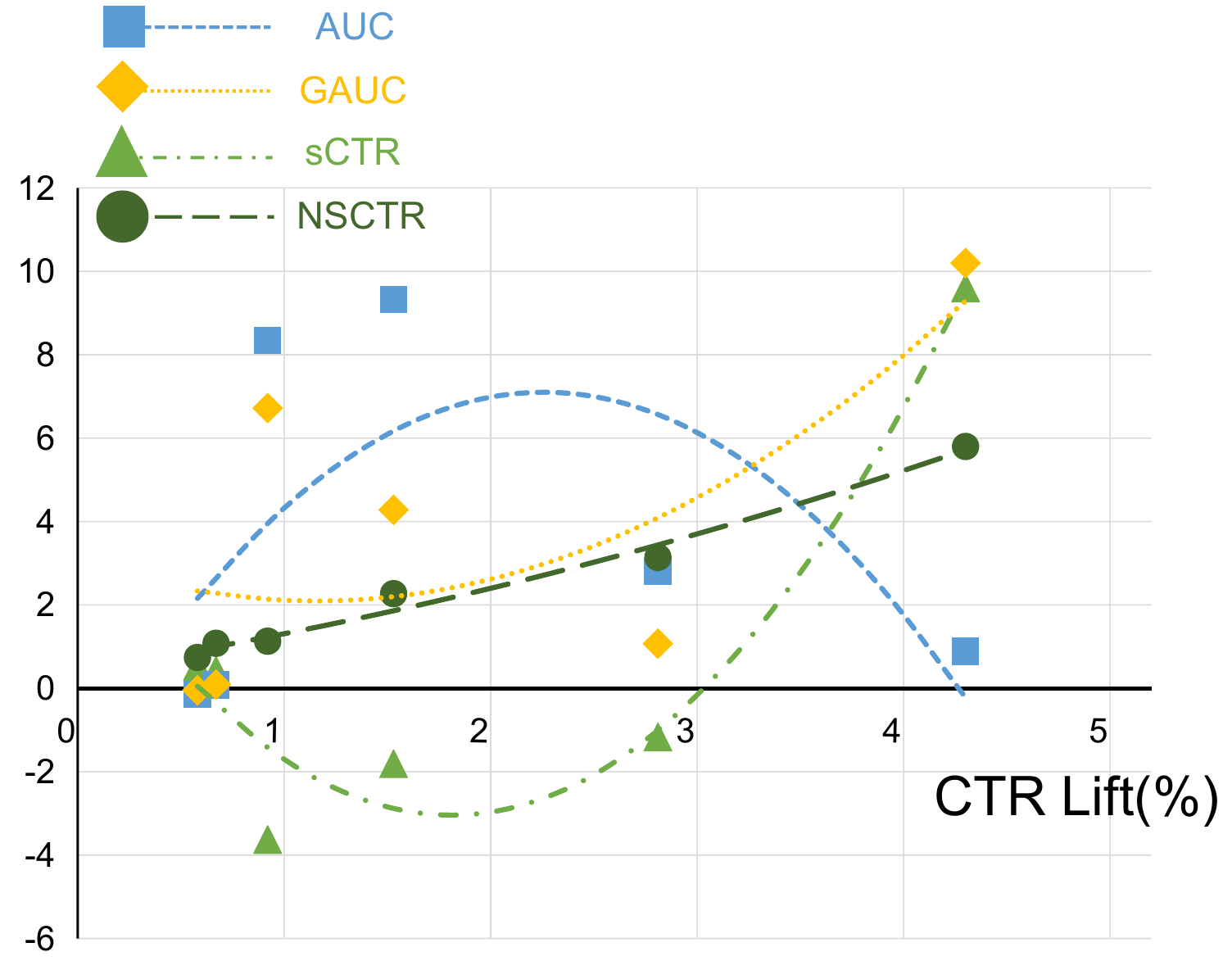}
\caption{The results between offline metrics and online A/B CTR lift for 6 major creative ranking upgrades. 
The x-axis shows online CTR lift in A/B tests. 
The y-axis shows the lift on 4 offline evaluation metrics. 
Each of the 6 discrete data points represents one major upgrade. 
Dashed lines show the quadratic trend between online and offline metrics.}
\label{fig4}
\end{figure}

Validating the effectiveness of offline evaluation metrics is challenging. 
The definitive test of efficacy lies in online A/B testing, but this approach often demands considerable time and resources. 
This motivates the necessity for offline metrics, which can be validated by correlating with online A/B results. 
To demonstrate this, we analyzed 6 major creative ranking upgrades over the past year, including bandit algorithms, personalized two-towers, creative feature extraction, etc. 
We computed 4 offline metric scores(AUC and GAUC measure the absolute lift at percentile points, while sCTR and NSCTR calculate the relative lift.) and corresponding online CTR lifts for each upgrade. 
The Pearson correlation coefficients were 0.988 for NSCTR, 0.636 for sCTR, 0.741 for AUC, and even -0.152 for GAUC. 
Plotting these in Fig.\ref{fig4} and calculating correlations with CTR shows that the NSCTR is a more reliable metric than others in the offline evaluation of creative ranking. 

\section{Experiments}
In this section, we conduct extensive experiments to evaluate the
effectiveness and efficiency of the proposed framework. 

\subsection{Implementation Details}
The AR model adopts rich features and complex network architectures, including more than 30 user features, more than 40 ad features, more than 300 user-ad cross features, as well as behavior sequences of users' historical information and ads. 
All the features are looked up into an embedding table of size $2^{30}$ with embedding dimension 16. 
On the other hand, a 2-layer transformer with 2 attention heads is used to model the user behavior sequences. 
The DCN has 4 hidden layers with sizes of 512 × 512 × 256 × 128. 

In contrast, the CR model uses simpler features and network structures, including 11 user features, 5 ad features, creative ID features, and content features. 
The embedding dimension is 4. 
The MLP for CR has 3 hidden layers with sizes of 128 × 64 × 32. 
The $pctr_{ad}$ embedding size $K$ is set as 8192 and the dimension $D$ is 128 by default. 

Both models use ReLU activation and sigmoid output layer to bound predictions within (0, 1). 
The batch size is 512 with Adagrad optimizer at a 0.05 learning rate. 
As usual in ranking models, the training epoch is set to 1, and we did not use the dropout strategy.

\subsection{Dataset}
Experiments are conducted on a log dataset gathered from a real-world ad system from May 1st to June 30th. 
We use the data of the first 60 consecutive days as the training set and that of June 30th as the test set. 
In total, There were about 18 billion training samples and 300 million test samples. 
The test set covers 53 million users, 16 million ads, and 54 million ad creatives. 
The feedback data for creatives is very sparse, with each creative receiving an average of only 6 impressions. 
The overall CTR of the samples is 2.4\%. 
To preserve the user feedback information as much as possible, we did not perform negative sampling on the training and test data. 
We evaluate our method in both offline and online settings to justify its design and performance. 

\subsection{Evaluation metrics} 
To evaluate the effectiveness of our proposed creative selection model, we adopt distinct evaluation metrics for offline and online experiments respectively. 

For offline experiments, we use Simulated Click-Through Rate (sCTR) and NSCTR to evaluate creative ranking performance. 
While we utilize Area Under ROC(AUC), and Group AUC(GAUC) as evaluation metrics in ad ranking. 

For online experiments, Click-Through Rate(CTR) and response time(RT) serve as the primary metrics. 
CTR measures the ratio of users clicking on an ad to total viewers and is widely used in online advertising. 
RT is another critical metric, as the system must return results within a certain time budget. 
Beyond CTR, we evaluate our model using Revenue per Mille(RPM), aligning with the goal of improving advertising revenue for the platform. 

\subsection{Baselines}
We implemented two categories of state-of-the-art methods as well as a naive baseline without creative ranking:
\begin{itemize}
    \item no-CR: no creative ranking module, each ad displays random candidate creatives, ad ranking module takes creative features as input.
    \item Post-CR: creative module cascaded after ad ranking module, the creative ranking module uses the two-tower model same as DSSM\cite{dssm:13}, ad ranking module does not use creative features for estimation. 
    Based on Post-CR, we update the creative ranking module with the CR+ model guided by our proposed JAC, forming a stronger baseline called Post-CR+.
    \item Pre-CR(CACS)\cite{CACS:22}: creative module cascaded before ad ranking module, ad ranking module uses creative features for estimation, creative ranking uses the two-tower model.
\end{itemize}

Additionally, we conduct ablation studies to analyze the effects of our proposed offline JAC model on creative ranking and ad ranking tasks separately. 
Peri-CR denotes the creative module in parallel with ad ranking module, but creative model uses an independently trained standalone MLP network. 
CR+ and AR+ are denoted as the creative and ad sub-module in JAC. 

\subsection{Hypotheses}
We expected Peri-CR would outperform Pre-CR and Post-CR in terms of response time, CTR, and CPM. 
We also expected JAC would achieve the best performance in both ad and creative ranking. 
Moreover, we expected JAC can help AR+ and CR+ perform separately. 

\begin{itemize}
    \item \textit{H1}: Peri-CR would be more efficient, i.e., less response time, than Pre-CR and Post-CR. 
    \item \textit{H2}: Peri-CR would be more effective, i.e., more CTR and CPM, than Pre-CR and Post-CR. 
    \item \textit{H3}: JAC would be most effective in both ad and creative ranking, i.e., achieving higher auc and gauc than AR, and higher sctr than CR. 
    \item \textit{H4}: JAC would help CR+ improve effectiveness, i.e., more sctr than CR without increasing time. 
    \item \textit{H5}: JAC would help AR+ improve effectiveness, i.e., more auc and gauc than AR without increasing time. 
\end{itemize}

\subsection{Online Evaluations}

\begin{table}[]
\caption{Performance of online architecture in online A/B test in 5 consecutive days. }
\label{tab:online1}
\begin{tabular}{cccc}
\hline
Method       & CTR               & RPM              & RT(ms)           \\ \hline
no-CR        & -                 & -                & -\textbf{(90ms)} \\
Post-CR      & +6.25\%           & +5.63\%          & +4.44\%(94ms)           \\
Pre-CR & +8.54\%           & +6.81\%          & +18.9\%(107ms)          \\
Peri-CR+     & \textbf{+10.12\%} & \textbf{+7.67\%} & -\textbf{(90ms)}  \\ \hline
\end{tabular}
\end{table}

\begin{table}[]
\caption{Online A/B Testing Results in 5 consecutive days with and without JAC. }
\label{tab:online2}
\begin{tabular}{cccc}
\hline
Method       & CTR               & RPM              & RT           \\ \hline
Post-CR(DSSM)      & -           & -          & -(94ms)           \\
Post-CR+     & +3.73\%  & +1.96\% & +4.17\%(98ms)           \\\hline
Peri-CR      & -           & -          & -(90ms)  \\
Peri-CR+     & +0.92\% & +0.52\% & -(90ms)  \\ \hline
AR          &   -   &   -   &   -(90ms) \\
AR+         &   +0.78\% &   +0.55\% &   -(90ms) \\ \hline
\end{tabular}
\end{table}

To verify the effectiveness of our proposed Peri-CR architecture on the overall ad system, we evaluated two key criteria - CTR and RPM.
As presented in Table.\ref{tab:online1}:
\begin{itemize}
    \item no-CR demonstrates significantly worst performance, indicating creative ranking can effectively select creatives that users are interested in to increase platform revenue. 
    \item Peri-CR improves CTR by 1.58\% and RPM by 0.86\% compared to Pre-CR, which confirms the hypothesis H2.
    \item Peri-CR+ shows similar performance as expected compared to Post-CR+. 
    \item Both Peri-CR+ over Peri-CR, Post-CR+ over Post-CR, and AR+ over AR have significant gains, validating that the offline JAC teacher model can effectively improve the online estimation performance. This result confirms the hypothesis H4 and H5. 
\end{itemize}

The response time(RT) of the ranking stage is the most important criterion for evaluating system efficiency. Our key observations are: 
\begin{itemize}
    \item Peri-CR+ exhibits the lowest response time (RT) comparable to no-CR, validating that our proposed architecture does not increase system latency. 
    This result confirms the hypothesis H1. 
    \item The marginal differences between Peri-CR+ vs Peri-CR and Post-CR+ vs Post-CR show offline JAC guidance introduces negligible computational overhead. 
    This further verifies hypothesis H4.
    \item Pre-CR \textgreater Post-CR \textgreater no-CR matches expectations, indicating more complex creative ranking under current architectures leads to higher latency. 
    \item It is worth noting that the CR in Peri-CR currently uses 28ms, sharing a 90ms time budget with AR. 
    This means we can further improve performance by adopting more complex feature extraction and model structures without increasing overall system latency.
\end{itemize}

In conclusion, Peri-CR achieves optimal performance in both system efficiency and effectiveness.

\subsection{Offline Evalutions}

Due to performance constraints within our online systems, we took an offline experimental approach to exploring the application potential of the JAC model. 

\begin{table}[]
\centering
\caption{Offline evaluation results for creative ranking. }
\label{tab:offline1}
\begin{tabular}{ccc}
\hline
        & sCTR & NSCTR \\ \hline
VAM-HBM & 0.2528    &  0.2267      \\
CACS    & 0.2589(+2.4\%)    &  0.2319(+2.3\%)      \\
CR      & 0.2559(+1.2\%)    &  0.2392(+5.5\%)      \\
CR+     & 0.2492(-1.4\%)    &  \textbf{0.2423(+6.9\%)}      \\ \hline
JAC     & 0.2478(-2.0\%)    &  \textbf{0.2427(+7.0\%)}      \\ \hline
\end{tabular}
\end{table}

Firstly, we validated the model's effectiveness at creative ranking. 
Results are demonstrated in Table.\ref{tab:offline1}. 
\begin{itemize}
    \item Compared to CR, CACS, and VAM-HBM, we can see that personalized creative ranking, as well as more complex network modeling, can significantly improve creative performance. 
    \item JAC achieved the highest NSCTR score, demonstrating that our offline joint modeling has superior creative ranking capabilities, which confirms hypothesis H3. 
    \item CR+ improved NSCTR by 1.2\% over CR, and approached the score of JAC, validating the effectiveness of our JAC network design in transferring the creative ranking capabilities of large models to smaller ones. 
    This further confirms the hypothesis H4. 
\end{itemize}

\begin{table}[]
\centering
\caption{Performance of offline evaluation for ad ranking. }
\label{tab:offline2}
\begin{tabular}{ccc}
\hline
    &   AUC      &    GAUC \\ \hline
AR  & 74.58\%    &   71.83\%   \\
AR+ & 74.70\%(+0.12\%)    &   71.98\%(+0.15\%)   \\ \hline
ACR & \textbf{74.87\%(+0.29\%)}    &   \textbf{72.14\%(+0.31\%)}   \\ \hline
\end{tabular}
\end{table}

Subsequently, we continued our discussion by investigating the JAC model’s potential strengths for ad ranking. 
As shown in Table.\ref{tab:offline2}, JAC also significantly outperforms on AUC and GAUC, demonstrating that creative features can help improve click-through rate prediction accuracy for ads. 
Hypothesis H3 hereof is confirmed. 
AR+ outperforms the baseline AR, validating that our offline large model can also help the small model better estimate ad click-through rates. 
This further confirms the hypothesis H5. 

\section{Limitations and Future Work}
Despite extensive online and offline experiments demonstrating the efficacy of our method, there are several limitations and future research directions: 

\begin{itemize}
    \item Due to latency and computation constraints online, we had to forgo a large joint ranker for ads and creatives, limiting optimal system performance. 
    With improving hardware and more efficient system design, we believe the joint ranker can achieve better performance without drastic latency increases.
    \item For offline metrics, NSCTR better correlates with online metrics. But it only approximates the true sample distribution to a certain degree. 
    We would like to find an instrumental variable in causal inference to perfectly recover the real distribution. 
    \item Last but not least, our current system separates offline creative production and online estimation. 
    With progress in AIGC, future systems could integrate them, generating optimal creatives in real-time tailored to each user.
\end{itemize}

\section{Conclusion}
In conclusion, this work makes several key contributions to improving ad and creative ranking in online advertising. 
First, we propose a novel architecture for online parallel estimation of ad and creative ranking, enabling sophisticated personalized creative modeling while reducing overall latency compared to conventional serial ranking approaches. 
Second, an offline joint model is constructed to allow mutual awareness and collaborative optimization between ads and creatives in CTR estimation, leading to improved accuracy. 
Third, we optimize the offline evaluation metrics for the implicit feedback sorting task critical for creative ranking, enhancing the offline-to-online correlation.
Extensive experiments demonstrate the effectiveness and efficiency of our approach over state-of-the-art methods, in both offline evaluations and real-world online advertising platforms. 
Specifically, our method achieves superior performance in terms of response time, CTR, and CPM compared to serial ranking approaches. 

\section{Acknowledgments}
The work was supported by the Creative Selection Team of JD.com's Recommendation Advertising Group. Special thanks to Junjie Li, Wei Hong, Xinyao Sun, and Zhuoya Yang for their contributions to the project. 

\bibliography{aaai24}

\begin{thebibliography}{30}
\providecommand{\natexlab}[1]{#1}

\bibitem[{Baltescu et~al.(2022)Baltescu, Chen, Pancha, Zhai, Leskovec, and Rosenberg}]{ItemSage:22}
Baltescu, P.; Chen, H.; Pancha, N.; Zhai, A.; Leskovec, J.; and Rosenberg, C. 2022.
\newblock ItemSage: Learning Product Embeddings for Shopping Recommendations at Pinterest.
\newblock In \emph{Proceedings of the 28th ACM SIGKDD Conference on Knowledge Discovery and Data Mining}, 2703–2711. New York, NY, USA.

\bibitem[{Brooks and Kugler(1987)}]{silver}
Brooks, F.; and Kugler, H. 1987.
\newblock \emph{No silver bullet}.
\newblock April.

\bibitem[{Brown et~al.(2020)Brown, Mann, Ryder, Subbiah, Kaplan, Dhariwal, Neelakantan, Shyam, Sastry, Askell, Agarwal, Herbert-Voss, Krueger, Henighan, Child, Ramesh, Ziegler, Wu, Winter, Hesse, Chen, Sigler, Litwin, Gray, Chess, Clark, Berner, McCandlish, Radford, Sutskever, and Amodei}]{gpt3:20}
Brown, T.; Mann, B.; Ryder, N.; Subbiah, M.; Kaplan, J.~D.; Dhariwal, P.; Neelakantan, A.; Shyam, P.; Sastry, G.; Askell, A.; Agarwal, S.; Herbert-Voss, A.; Krueger, G.; Henighan, T.; Child, R.; Ramesh, A.; Ziegler, D.; Wu, J.; Winter, C.; Hesse, C.; Chen, M.; Sigler, E.; Litwin, M.; Gray, S.; Chess, B.; Clark, J.; Berner, C.; McCandlish, S.; Radford, A.; Sutskever, I.; and Amodei, D. 2020.
\newblock Language Models are Few-Shot Learners.
\newblock In \emph{Advances in Neural Information Processing Systems}, volume~33, 1877--1901.

\bibitem[{Chen et~al.(2021{\natexlab{a}})Chen, Ge, Jiang, Zhang, Lian, and Zheng}]{chen2021efficient}
Chen, J.; Ge, T.; Jiang, G.; Zhang, Z.; Lian, D.; and Zheng, K. 2021{\natexlab{a}}.
\newblock Efficient Optimal Selection for Composited Advertising Creatives with Tree Structure.
\newblock In \emph{Proceedings of the AAAI Conference on Artificial Intelligence}, volume~35, 3967--3975.

\bibitem[{Chen et~al.(2021{\natexlab{b}})Chen, Xu, Jiang, Ge, Zhang, Lian, and Zheng}]{chen2021automated}
Chen, J.; Xu, J.; Jiang, G.; Ge, T.; Zhang, Z.; Lian, D.; and Zheng, K. 2021{\natexlab{b}}.
\newblock Automated Creative Optimization for E-Commerce Advertising.
\newblock In \emph{Proceedings of the Web Conference 2021}, 2304--2313.

\bibitem[{Cheng et~al.(2016)Cheng, Koc, Harmsen, Shaked, Chandra, Aradhye, Anderson, Corrado, Chai, Ispir et~al.}]{wd:16}
Cheng, H.-T.; Koc, L.; Harmsen, J.; Shaked, T.; Chandra, T.; Aradhye, H.; Anderson, G.; Corrado, G.; Chai, W.; Ispir, M.; et~al. 2016.
\newblock Wide \& deep learning for recommender systems.
\newblock In \emph{Proceedings of the 1st workshop on deep learning for recommender systems}, 7--10.

\bibitem[{Dang et~al.(2023)Dang, Yang, Guo, Jiang, Wang, Xu, Sun, and Liu}]{uniform:23}
Dang, Y.; Yang, E.; Guo, G.; Jiang, L.; Wang, X.; Xu, X.; Sun, Q.; and Liu, H. 2023.
\newblock Uniform Sequence Better: Time Interval Aware Data Augmentation for Sequential Recommendation.
\newblock In \emph{Proceedings of the AAAI Conference on Artificial Intelligence}, volume~37, 4225--4232.

\bibitem[{Fan et~al.(2019)Fan, Guo, Zhu, Miao, Sun, and Li}]{mobius:19}
Fan, M.; Guo, J.; Zhu, S.; Miao, S.; Sun, M.; and Li, P. 2019.
\newblock MOBIUS: Towards the Next Generation of Query-Ad Matching in Baidu's Sponsored Search.
\newblock In \emph{Proceedings of the 25th ACM SIGKDD International Conference on Knowledge Discovery \& Data Mining}, 2509–2517. New York, NY, USA.

\bibitem[{Fawcett(2006)}]{ROC:06}
Fawcett, T. 2006.
\newblock An introduction to ROC analysis.
\newblock \emph{Pattern recognition letters}, 27(8): 861--874.

\bibitem[{Halinen(1996)}]{mkt:96}
Halinen, A. 1996.
\newblock \emph{Relationship Marketing in Professional Services: A Study of Agency-Client Dynamics in the Advertising Sector}.
\newblock London: Routledge.

\bibitem[{Huang et~al.(2013)Huang, He, Gao, Deng, Acero, and Heck}]{dssm:13}
Huang, P.-S.; He, X.; Gao, J.; Deng, L.; Acero, A.; and Heck, L. 2013.
\newblock Learning Deep Structured Semantic Models for Web Search Using Clickthrough Data.
\newblock In \emph{Proceedings of the 22nd ACM International Conference on Information \& Knowledge Management}, 2333–2338. New York, NY, USA.

\bibitem[{Jin et~al.(2021)Jin, Tiezheng, Gangwei, Zhiqiang, Defu, and Kai}]{aes:21}
Jin, C.; Tiezheng, G.; Gangwei, J.; Zhiqiang, Z.; Defu, L.; and Kai, Z. 2021.
\newblock Efficient Optimal Selection for Composited Advertising Creatives with Tree Structure.
\newblock In \emph{Proceedings of the AAAI Conference on Artificial Intelligence}, volume~35, 3967--3975.

\bibitem[{Li et~al.(2023)Li, Wang, Li, Fu, Shen, Shang, and McAuley}]{textseq:23}
Li, J.; Wang, M.; Li, J.; Fu, J.; Shen, X.; Shang, J.; and McAuley, J. 2023.
\newblock Text Is All You Need: Learning Language Representations for Sequential Recommendation.
\newblock In \emph{Proceedings of the 29th ACM SIGKDD Conference on Knowledge Discovery and Data Mining}, 1258–1267. New York, NY, USA.

\bibitem[{Li et~al.(2019)Li, Chen, Pettit, and Rijke}]{rerank:19}
Li, X.; Chen, Y.; Pettit, B.; and Rijke, M.~D. 2019.
\newblock Personalised Reranking of Paper Recommendations Using Paper Content and User Behavior.
\newblock \emph{ACM Trans. Inf. Syst.}, 37(3).

\bibitem[{Lin et~al.(2022)Lin, Zhang, Li, Wang, Long, Deng, Xu, and Zheng}]{CACS:22}
Lin, K.; Zhang, X.; Li, F.; Wang, P.; Long, Q.; Deng, H.; Xu, J.; and Zheng, B. 2022.
\newblock Joint Optimization of Ad Ranking and Creative Selection.
\newblock In \emph{Proceedings of the 45th International ACM SIGIR Conference on Research and Development in Information Retrieval}, 2341–2346. New York, NY, USA.

\bibitem[{Liu et~al.(2017)Liu, Xiao, Ou, and Si}]{cascade:17}
Liu, S.; Xiao, F.; Ou, W.; and Si, L. 2017.
\newblock Cascade Ranking for Operational E-Commerce Search.
\newblock In \emph{Proceedings of the 23rd ACM SIGKDD International Conference on Knowledge Discovery and Data Mining}, 1557–1565. New York, NY, USA.

\bibitem[{Rombach et~al.(2022)Rombach, Blattmann, Lorenz, Esser, and Ommer}]{sd:22}
Rombach, R.; Blattmann, A.; Lorenz, D.; Esser, P.; and Ommer, B. 2022.
\newblock High-Resolution Image Synthesis With Latent Diffusion Models.
\newblock In \emph{Proceedings of the IEEE/CVF Conference on Computer Vision and Pattern Recognition (CVPR)}, 10684--10695.

\bibitem[{Shiwei et~al.(2023)Shiwei, Huifeng, Lu, Wei, Xing, Ruiming, Rui, and Ruixuan}]{alt:2023}
Shiwei, L.; Huifeng, G.; Lu, H.; Wei, Z.; Xing, T.; Ruiming, T.; Rui, Z.; and Ruixuan, L. 2023.
\newblock Adaptive Low-Precision Training for Embeddings in Click-Through Rate Prediction.
\newblock In \emph{Proceedings of the AAAI Conference on Artificial Intelligence}, 4435--4443.

\bibitem[{Talebi and Milanfar(2018)}]{nima:18}
Talebi, H.; and Milanfar, P. 2018.
\newblock NIMA: Neural image assessment.
\newblock \emph{IEEE transactions on image processing}, 27(8): 3998--4011.

\bibitem[{Van Den~Oord, Vinyals, and Kavukcuoglu(2017)}]{VQVAE:17}
Van Den~Oord, A.; Vinyals, O.; and Kavukcuoglu, K. 2017.
\newblock Neural discrete representation learning.
\newblock In \emph{Advances in the 30th Neural Information Processing Systems}, volume~30, 6309–6318.

\bibitem[{Wang et~al.(2017)Wang, Fu, Fu, and Wang}]{dcn:17}
Wang, R.; Fu, B.; Fu, G.; and Wang, M. 2017.
\newblock Deep \& Cross Network for Ad Click Predictions.
\newblock In \emph{Proceedings of the ADKDD'17}, 1--7. New York, NY, USA.

\bibitem[{Wang et~al.(2021)Wang, Liu, Ge, Lian, and Zhang}]{HBM:21}
Wang, S.; Liu, Q.; Ge, T.; Lian, D.; and Zhang, Z. 2021.
\newblock A Hybrid Bandit Model with Visual Priors for Creative Ranking in Display Advertising.
\newblock In \emph{Proceedings of the Web Conference 2021}, 2324–2334. New York, NY, USA.

\bibitem[{Wang et~al.(2022)Wang, Zhao, Xu, and Wu}]{auto:22}
Wang, Y.; Zhao, X.; Xu, T.; and Wu, X. 2022.
\newblock Autofield: Automating feature selection in deep recommender systems.
\newblock In \emph{Proceedings of the ACM Web Conference 2022}, 1977--1986.

\bibitem[{Xi et~al.(2021)Xi, Liu, Dai, Tang, Zhang, Liu, He, and Yu}]{rerank:21}
Xi, Y.; Liu, W.; Dai, X.; Tang, R.; Zhang, W.; Liu, Q.; He, X.; and Yu, Y. 2021.
\newblock Context-aware reranking with utility maximization for recommendation.
\newblock \emph{arXiv preprint arXiv:2110.09059}.

\bibitem[{Xiao et~al.(2023)Xiao, Bo, Chenrui, Han, and Mingchen}]{gauc}
Xiao, S.; Bo, Z.; Chenrui, Z.; Han, R.; and Mingchen, C. 2023.
\newblock Enhancing Personalized Ranking With Differentiable Group AUC Optimization.
\newblock arXiv:2304.09176.

\bibitem[{Yang et~al.(2023)Yang, Chen, Yu, Wu, Ma, Zhao, Fang, Chen, Fan, He et~al.}]{yang2023incremental}
Yang, C.; Chen, J.; Yu, Q.; Wu, X.; Ma, K.; Zhao, Z.; Fang, Z.; Chen, W.; Fan, C.; He, J.; et~al. 2023.
\newblock An Incremental Update Framework for Online Recommenders with Data-Driven Prior.
\newblock In \emph{Proceedings of the 32nd ACM International Conference on Information and Knowledge Management}, 4894--4900.

\bibitem[{Zhang et~al.(2023)Zhang, Huang, Ou, Li, Li, Liu, and Zeng}]{prerank:23}
Zhang, Z.; Huang, Y.; Ou, D.; Li, S.; Li, L.; Liu, Q.; and Zeng, X. 2023.
\newblock Rethinking the Role of Pre-ranking in Large-scale E-Commerce Searching System.
\newblock arXiv:2305.13647.

\bibitem[{Zhao et~al.(2019)Zhao, Li, Zhang, Wang, Jiang, Xu, Wang, and Ma}]{peac:19}
Zhao, Z.; Li, L.; Zhang, B.; Wang, M.; Jiang, Y.; Xu, L.; Wang, F.; and Ma, W. 2019.
\newblock What You Look Matters? Offline Evaluation of Advertising Creatives for Cold-Start Problem.
\newblock In \emph{Proceedings of the 28th ACM International Conference on Information and Knowledge Management}, 2605–2613. New York, NY, USA.

\bibitem[{Zhou et~al.(2019)Zhou, Mou, Fan, Pi, Bian, Zhou, Zhu, and Gai}]{dien:19}
Zhou, G.; Mou, N.; Fan, Y.; Pi, Q.; Bian, W.; Zhou, C.; Zhu, X.; and Gai, K. 2019.
\newblock Deep Interest Evolution Network for Click-through Rate Prediction.
\newblock In \emph{Proceedings of the AAAI Conference on Artificial Intelligence}.

\bibitem[{Zhou et~al.(2018)Zhou, Zhu, Song, Fan, Zhu, Ma, Yan, Jin, Li, and Gai}]{din:18}
Zhou, G.; Zhu, X.; Song, C.; Fan, Y.; Zhu, H.; Ma, X.; Yan, Y.; Jin, J.; Li, H.; and Gai, K. 2018.
\newblock Deep Interest Network for Click-Through Rate Prediction.
\newblock In \emph{Proceedings of the 24th ACM SIGKDD International Conference on Knowledge Discovery and Data Mining}, 1059–1068. New York, NY, USA.

\end{thebibliography}

\end{document}